\documentclass[12pt]{article}

\usepackage{epsfig}
\usepackage{amsmath}
\usepackage{multirow}

\newcommand{\al}{A^{{}^L}}
\newcommand{\ah}{A^{{}^H}}
\newcommand{\ms}{m_{\scriptscriptstyle SUSY}}
\newcommand{\bmssm}{b^{\scriptscriptstyle MSSM}}
\newcommand{\bsm}{b^{\scriptscriptstyle SM}}

\begin{document}


\begin{flushright} CERN--TH/2001--216 \end{flushright}
\vspace{5mm}
\vspace{0.5cm}
\begin{center}

\def\thefootnote{\fnsymbol{footnote}}

{\Large \bf Running and Matching from 5 to 4 Dimensions} \\[1cm]
{\large R. Contino$^1$, L. Pilo$^1$, R. Rattazzi$^{2\,}$\footnote{On leave 
 from INFN, Pisa, Italy.}, E. Trincherini$^3$}
\\[1.5cm]

{\small 
$^1$\textit{Scuola Normale Superiore, Piazza dei Cavalieri 7, I-56126 Pisa,
Italy $\&$ INFN} 
\\[0.3cm] $^2$\textit{Theory division, CERN, CH-1211 Geneva 23, Switzerland}
\\[0.3cm]
$^3$\textit{Physics Department, University of Pisa $\&$ INFN, sez. Pisa,
I-56126 Pisa, Italy} }

\end{center}

\vspace{1cm}

\hrule \vspace{0.3cm} 
{\small  \noindent \textbf{Abstract} \\[0.3cm]
\noindent
We study 5 dimensional grand-unified theories in an orbifold geometry by the method of 
effective field theory: we match the low and high energy theories by integrating
out at \mbox{1-loop} the massive Kaluza-Klein states.
In the supersymmetric case the radius dependence of threshold effects 
is fixed by the rescaling anomalies of the low energy theory. 
We focus on 
a recently proposed $SU(5)$  model on $M^4  \times S^1/(Z_2\times Z_2^\prime)$.
Even though the spectrum of the heavy modes is completely known, there still are
corrections to gauge unification originating from boundary couplings. 
In order to control these effects one has to rely
on extra assumptions. We argue that, as far as gauge couplings are concerned,
the predictive power of these models is similar to conventional GUTs.}
\vspace{0.5cm}  \hrule

\def\thefootnote{\arabic{footnote}}
\setcounter{footnote}{0}


\section{Introduction}

Since their early days \cite{GG} grand unified theories (GUT) have attracted a 
lot of interest and though the paradigm is not flawless it is still an 
exciting and active research field. Besides aesthetic reasons, a strong 
evidence in favor of GUTs is that in the minimal supersymmetric standard model
(MSSM) the couplings of $SU(3)\times SU(2)\times U(1)$ unify at a scale of
order $10^{16}$ GeV \cite{Lan}.
Models with extra compact dimensions can add interesting twists to
the basic GUT idea. This was first realized in string models
where intrinsically higher dimensional mechanisms 
 can solve some of the problems of conventional GUTs. One example is the
doublet-triplet splitting problem \cite{witten85}. Recently there has been a revival
in extra-dimensional GUT model building, but now taking a ``bottom-up'' approach
as opposed to the ``top-down'' approach of string model building.
Kawamura has first  constructed a non-supersymmetric model on $S^1/Z_2$ \cite{kawa},
and has later obtained a realistic supersymmetric spectrum  
on $S^1/(Z_2\times Z_2^\prime)$ \cite{kawa2}. (The interesting properties of 
$S^1/(Z_2\times Z_2^\prime)$ for model building where noticed in ref. \cite{BHN}).
The model was further studied in \cite{af,HN}.
Many papers have since followed \cite{all}-\cite{BHN1}.
In this paper we extend the effective field theory (EFT) approach proposed by 
Weinberg \cite{weinberg} to the case of a grand unified gauge theory in five 
dimensions. We build the effective field theory 
valid below the GUT scale by integrating out  the heavy degrees of freedom
represented here by the massive KK excitations. The form of the low-energy 
theory is strongly constrained by symmetries and operators' coefficients
are expressed in terms of the parameters of the underlying 
high energy theory. As an explicit example we consider the $SU(5)$ unified 
theory on $M^4  \times S^1/(Z_2\times Z_2^\prime)$ of  
ref.~\cite{kawa2,HN}. We compute the matching functions relating the SM gauge 
couplings to the parameters of the 5 dimensional theory.
Although the explicit results are given for $SU(5)$, the method presented  
is general. 
One  important feature of the orbifold geometry is the presence at the boundary of 
local operators contributing to the vector boson kinetic term. These operators
do not respect $SU(5)$ and lead to corrections to gauge unification. 
Indeed their coefficients follow a logarithmic
RG evolution, so that they cannot just be set to zero.

The outline of the paper is the following. In section 2 we discuss the 
conditions under which the gauge symmetry does not conflict with the orbifold
projections and the $SU(5)$ model considered is briefly reviewed.
Section 3 is devoted to a general discussion of the boundary counterterms 
induced by quantum corrections at the orbifold fixed points. The construction 
of the effective 4D low energy theory is presented in section 4, in particular
the matching functions are computed at the one loop level.
Finally, in section  5, the values of the SM 
gauge couplings are computed at the weak scale at next to leading order 
and the phenomenological consequences are discussed.

\section{$SU(5)$ on the orbifold}

We consider a grand unified 5-dimensional $SU(5)$ theory defined on
$M^4\times S^1/(Z_2\times Z_2^\prime)$ \cite{kawa2,HN}, where $S^1/(Z_2\times 
Z_2^\prime)$ is obtained from a circle with radius $R$ by the following 
identifications:
\begin{equation}
Z_2 : \quad y \sim 2\pi R-y \, ,  \qquad 
Z_2^\prime  : \quad y \sim \pi R-y \, 
\qquad y\in [0,2\pi R].
\end{equation}
Coordinates in 5D  are denoted by $X^M=(x^\mu,y)$,
using capital Latin (small Greek) letters for 5D (4D) indices. 
The points $(0,\pi R)$ and $(\pi R/2, 3 \pi R/2 )$ are fixed under the action 
of $Z_2$ and  $Z_2^\prime$ respectively; moreover $0 \sim \pi R$ and $\pi R/2
\sim 3 \pi R/2$.

A function $f(x,y)$ with a definite parity under the orbifold projections
can be Fourier decomposed according to 
\begin{equation}
f(x,y) = \sum_{n=0}^{+\infty} f^{(n)}(x) \, \Psi_n(y), \qquad
\Psi_n(y) = \begin{cases} 
 a_{2n} \, \cos \displaystyle{\frac{2ny}{R}}  &  \quad (+,+) \\[0.25cm]
 a_{2n+1} \, \cos \displaystyle{\frac{(2n+1)y}{R}} &  \quad (+,-) \\[0.25cm]
 a_{2n+1} \, \sin \displaystyle{\frac{(2n+1)y}{R}} & \quad (-,+ ) \\[0.25cm]
 a_{2n+2} \, \sin \displaystyle{\frac{(2n+2)y}{R}}  &  \quad (-,-)  \end{cases} 
\end{equation}
\begin{equation*}
a_0 = \frac{1}{\sqrt{2\pi R}}, \quad  a_n = \frac{1}{\sqrt{\pi R}} \quad 
n\not=0 \, .
\end{equation*}
Only a function with parity $(+,+)$ has a zero mode.

When a gauge theory is defined in a orbifold one has to satisfy certain 
consistency conditions in order to avoid that gauge symmetry conflicts 
with orbifold projections.
The $Z_2$ action $P$ on the fields of the theory is defined to be
\begin{equation}
\Phi(x,y) \to \Phi(x,-y) = P \, \Phi(x,y) \qquad \qquad P^2 = 1 \; ,
\end{equation}
where we have collected all the fields in single vector $\Phi$.
Without loss of generality $P$ can be chosen diagonal with eigenvalues 
$\pm 1$. In particular, for the gauge field $A_M^i T^i$ one has
\begin{equation}
\begin{split}
A_\mu^i(x,-y) &= \Lambda^{ij} A_\mu^j(x,y) \\
A_5^i(x,-y) &= - \Lambda^{ij} A_5^j(x,y)
\end{split}
\qquad \qquad \Lambda^2 =1 \quad ;
\end{equation}
where $T^i$ are the generators of the gauge group $G$ with structure 
constants $f_{ijk}$. For the Lagrangian to be $Z_2$ invariant, 
the covariant derivative acting on a matter field $D_M \varphi$ must 
have a definite transformation property under $Z_2$, thus
\begin{equation}
P T^i P = \Lambda^{ij} T^j
\end{equation}
In turn this implies that $P$ is an inner automorphism of the Lie algebra
${\cal G}$ of $G$, in other words
\begin{equation} \label{eq:cons}
\Lambda_{im} \, \Lambda_{jn} \, \Lambda_{kl} \, f_{mnl} = f_{ijk} \, ,
\end{equation}
and $P$ acts as a group conjugation (see \cite{JMR} for a discussion of inner
and outer automorphism in the context of orbifold projections). 
Taking $\Lambda_{ij} = \delta_{ij} \, c_i$ (no summation in $i$), 
with $c_i = \pm 1$, we can divide the generators in two subsets: 
${\cal H} = \{T^a, \, a = 1, \, \cdots n \, : c_a = 1 \; \forall a \}$ and 
${\cal V} = \{T^{\hat a}, \, \hat{a} = n+1 , \, \cdots \text{dim}
({\cal G}) \, : c_{\hat{a}} = - 1 \; \forall \hat{a} \}$.
From eq.(\ref{eq:cons}) it follows that $[{\cal H}, {\cal H}] = {\cal H}$,
$[{\cal H}, {\cal V}] = {\cal V}$ and $[{\cal V}, {\cal V}] = {\cal H}$.
Repeating the same argument, with the simplifying hypothesis that also 
second projection $Z_2^\prime$ acts through $P^\prime$ as a diagonal matrix 
on the fields, exactly the same result holds~\footnote{For a discussion
of the general case see \cite{BHN1}.}.
Summarizing, consistency between the gauge symmetry and orbifold projections 
leads to a $Z_2$ gradation for ${\cal G}$
\begin{equation}
\begin{gathered}
\, [{\cal H}, {\cal H}] = {\cal H} \,,  \quad [{\cal H}, {\cal V}] = {\cal V} \, ,
\quad [{\cal V}, {\cal V}] = {\cal H} \quad ; \\
\forall \, T^a \in {\cal H} \text{ with parity  } (+,+) \quad \text{and }
\forall \, T^{\hat{a}} \in {\cal V} \text{ with parity  } (-,\cdot \, ) 
\text{ or }  (\cdot\, ,-) \quad .
\end{gathered}
\end{equation}
Finally, the parity of a generic field must be gauge independent, 
implying that a gauge transformation with parameters $\xi^i(x,y)$
commutes with the $Z_2\times Z_2^\prime$ action
and as a consequence $\xi^i(x,y)$ must have the same parity of the 
corresponding $A_\mu^i$.

In \cite{kawa2,HN} a realistic supersymmetric $SU(5)$ unified theory in 5 dimensions 
was constructed with a vector gauge multiplet and two higgs hypermultiplets in 
the $5$ and $\bar 5$ representation propagating in the bulk. 
From the 4D point of view the field content is the one of N=2 SUSY. 
Indeed, the 5D vector multiplet splits into a vector and a chiral 
N=1 multiplets  $V=(A_\mu,\lambda)$, $\Sigma=(\sigma+i A_5,\lambda^\prime)$;
in the same way each of the hypermultiplets decomposes into two chiral N=1 
multiplets, $H,H^c$ and $\bar H,\bar H^c$ respectively.

The action of $P$ and $P^\prime$ is chosen to be \cite{kawa}
\begin{equation}
\begin{gathered}
P=\mathbf{1}_{5\times 5} \qquad 
P^\prime = \begin{pmatrix} \mathbf{1}_{3\times 3} & \\ & -\mathbf{1}_{2\times
2} 
\end{pmatrix} \quad .
\end{gathered}
\end{equation}
Denoting with $\{T^a, a=1,\dots 12 \}$  the $SU(3)\times SU(2)\times U(1)$ 
generators and with $\{T^{\hat a}, \hat a=13,\dots 24 \}$ the remaining ones,
one has a $(+,+)$ parity for $A^a_\mu$ and $(+,-)$ for $A^{\hat a}_\mu$.
With this choice, the zero modes gauge fields are the ones of the 
Standard Model. Notice that, while in the fixed point $O (y=0)$ the 
$SU(5)$ is still effective,
in $O^\prime (y=\pi R/2)$ only $SU(3)\times SU(2)\times U(1)$ gauge 
transformations are non-vanishing.
The parity of the various fields is summarized in table
\ref{tab:parity}.
\begin{table}
\begin{center}
\begin{tabular}{|c|c|c|}
\hline
 $(P,P')$ & 4D $N=1$ superfield & mass       \\ \hline & & \\[-0.25cm]
 $(+,+)$  & $V^a$, $H_2$, $\bar H_{2}$                  & $2n/R$     \\ 
 $(+,-)$  & $V^{\hat{a}}$, $H_3$, $\bar H_{3}$          & $(2n+1)/R$ \\ 
 $(-,+)$  & $\Sigma^{\hat{a}}$, $H_3^c$, $\bar H_{3}^c$ & $(2n+1)/R$ \\ 
 $(-,-)$  & $\Sigma^a$, $H_2^c$, $\bar H_{2}^c$         & $(2n+2)/R$ \\[0.2cm]
\hline
\end{tabular}
\end{center}
\caption{\textit{Parity assignments for the various fields. The subscripts
$2,3$ refer to the
weak doublet and the colour triplet of Higgs multiplets.}}
\label{tab:parity}
\end{table}
The choice is such that only the weak doublet components $H_2$, $\bar H_2$ of
the higgs multiplets $H$, $\bar H$ have zero modes, while the triplet components $H_3$,
$\bar H_3$ and the remaining two multiplets are massive.
It is also clear that the zero mode spectrum is N=1 supersymmetric.

Finally, the MSSM matter content, organized
in $SU(5)$ multiplets, and the Yukawa couplings are localized in the fixed 
point $O$ where the original gauge symmetry is unrestricted;
of course N=1 SUSY must be broken with conventional 4D methods to get an 
acceptable phenomenology. 
According to our classification, the other possible choice in 
which the generators 
$T^{\hat a}$ are $(-,-)$ would break $SU(5)$ in both fixed points rendering
unnatural the organization in $SU(5)$ multiplets of the observed matter.

\section{Boundary counterterms} \label{sect:boundary}

Before presenting the computation of the matching, it useful to 
gain some feeling on the general structure of the 1-loop radiative corrections of  a 
gauge theory on an orbifold.
The main difference with respect to the standard case is that, besides bulk counterterms,  
also operators localized at the orbifold fixed points can be induced~\cite{georgi}.

\begin{figure}
\centering
\epsfig{file=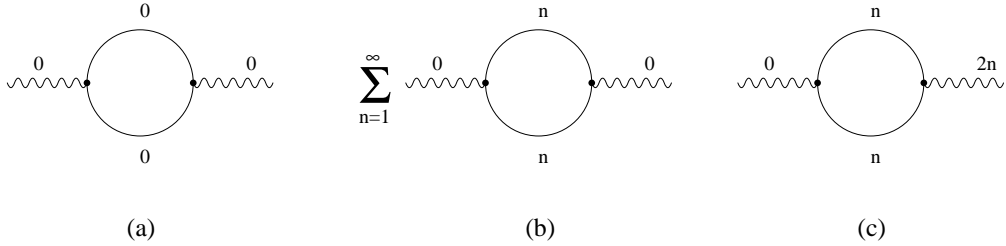,width=0.9\linewidth}
\caption{\textit{One-loop rainbow diagrams with at least a zero-mode external
line which contribute to the gauge kinetic term. Diagram (b) with all scalar, vector
and fermionic fields circulating in the loop is relevant for the computation 
of the matching functions. Diagram (c) is allowed only in the orbifold.}}
\label{fig:diagrams}
\end{figure}

We consider the 1-loop corrections to the gauge kinetic term due to a bulk scalar. 
For simplicity we focus on the diagrams with at least one zero-mode external line; the
rainbow-like Feynman graphs involved are shown in fig.(\ref{fig:diagrams}).
The contribution from seagull diagrams (with a quartic vertex)  
reproduces the transverse structure dictated by gauge invariance and 
will be neglected in our discussion.
The analysis can be easily extended to fermions circulating in the loop.
Some additional care would be needed in dealing with gauge bosons: the counterterms 
will be local provided the gauge fixing term is also local in 5D. 

Let us do some power counting. The gauge kinetic coefficient in 5D has dimension
$[{\rm mass}]$. Indicating by $\Lambda$ the UV cut-off, we in general expect for the bulk
kinetic term a divergence $\propto \Lambda +m \ln \Lambda$, where $m$ is any lagrangian mass
parameter (for instance a fermion mass). In our case there is no massive parameter in the bulk,
so we will not have any logarithmic divergence there. Moreover 
we choose to work with dimensional regularization \footnote{See for instance appendix~D
of ref. \cite{gravitons} for details on dimensional regularization 
in the presence of compact dimensions.}
(truly, dimensional reduction 
to preserve supersymmetry), for which there are also no power divergences 
\footnote{Dimensional regularization is useful in dealing with effective field 
theories \cite{weinberg} as it preserves only the logarithmic divergences, i.e.  those, and
only those, which are saturated in the infrared. Power divergences are totally UV dominated
and their effect is equivalent to changing the UV boundary conditions on the already 
incalculable parameters of the EFT. Note that as far as the logarithmic divergences are
concerned
our results agree with ref. \cite{ddg}, where gauge coupling evolution in 5 dimensions
was studied, but with a different regularization procedure.}.
So we expect no divergence whatsoever from the bulk. 
However our space is not homogeneous, it has fixed points, and it is perfectly fine
to have divergences located there. The fixed points are 4D manifolds, so here we 
will in general have logarithmic renormalization of the gauge kinetic term. Therefore the only
divergences are due to the boundaries.

Consider for instance the simple case of a scalar on a circle. Since there are no boundaries
the sum of diagrams (1a) and (1b) in 
fig.(\ref{fig:diagrams}) must be finite (diagram (1c) is not allowed on 
the circle).
A scalar on the circle can be decomposed into cosine and sine modes
both circulating in the loop of diagram (\ref{fig:diagrams}b).
The sum over cosine modes and the sum over the sine modes are equal.
Therefore their divergent piece should equal
$-1/2$ of the zero mode divergent contribution. This result can be checked by explicit
computation. Somehow the KK mode tower acts as a regulator of the UV divergences
of the zero mode. 

Consider next the non trivial case of an orbifold $S^1/Z_2$, with the  gauge 
field  taken with positive parity. Because of the orbifold projection, half of
the massive scalar modes (say the sines) are eliminated, 
the above cancellation no longer works and logarithmic divergences appear.
The only possible counterterm has the form
\begin{equation}
{\cal L}_{ct} = \int d^4x \, dy  \, F_{\mu\nu} F^{\mu\nu}  \big[a \,
\delta(y)+b \, \delta(y-\pi R)\big] \quad ,
\label{ct}
\end{equation}
where $a,b$ are constants; group indices have been suppressed for simplicity.
After a Fourier decomposition, (\ref{ct}) takes the form
\begin{equation}
\begin{split}
{\cal L}_{ct} = \int d^4x  \, \Big\{ 
 & (a+b) \sum_{n=0}^\infty  a_n^2 F_{\mu\nu}^{(n)} F^{\mu\nu\, (n)} + 
 2 (a+b) \sum_{\overset{\scriptstyle{n=0}}{k=1}}^\infty  a_n \, a_{n+2k} \,
  F_{\mu\nu}^{(n)} F^{\mu\nu\, (n+2k)} + \\
 & 2 (a-b) \sum_{\overset{\scriptstyle{n=0}}{k=1}}^\infty  a_n \, a_{n+2k+1} \,
  F_{\mu\nu}^{(n)} F^{\mu\nu\, (n+2k+1)} \Big\} \quad . 
\end{split}
\end{equation}
Momentum along the fifth dimension is conserved up to a sign and only 
``transitions'' $k\to (k+2n)$ are allowed requiring $a=b$. 
We have explicitly checked  that all the (logarithmic) divergences in diagrams 
(1a), (1b) and (1c) correspond to  $a=b$ for all parity choices of the 
scalar. If the gauge field has negative parity, no counterterm localized 
at the orbifold fixed points is possible and therefore the 1-loop correction 
to the gauge kinetic term must be finite in dimensional regularization.

Finally, we consider an $S^1/(Z_2\times Z_2^\prime)$ orbifold. The most general 
counterterm localized at the two (inequivalent) fixed points is 
\begin{equation}
{\cal L}_{ct} = \int d^4x \, dy  \, F_{\mu\nu} F^{\mu\nu}  \big[a \,
\delta(y)+b \, 
 \delta(y-\pi R/2)\big]
\label{ct2}
\end{equation}
According to the parity choice for $A_\mu^i$ and for the fields circulating in
the loop, we get different values for $a, b$ as shown in the following table.
\vspace{0.5cm}
\begin{center}
\begin{tabular}{c|cc|c}

 $ A_\mu^i$ & \multicolumn{2}{c|}{loop fields} & counterterm \\
\hline & & \\[-0.2cm]
 \multirow{2}{1.cm}{$(+,+)$} & $(\pm, \pm)$ & $(\pm, \pm)$ & $a=b$ \\[0.2cm]
                             & $(\pm, \mp)$ & $(\pm, \mp)$ & $a=-b$ \\[0.2cm]
 $(+,-)$ & $(\pm, \mp)$ & $(\pm, \pm)$ & $b=0$ \\[0.2cm]
 $(-,+)$ & $(\mp, \pm)$ & $(\pm, \pm)$ & $a=0$ \\[0.2cm]
\end{tabular}
\end{center}
\vspace{0.5cm}
Notice that $Z_2$ and $Z_2^\prime$  parities are separately conserved in each 
vertex and when the gauge field is odd (parity $(\pm, \mp)$) the only non 
vanishing diagrams are the ones with an even 
(parity ($\pm, \pm$)) and an odd field circulating in the loop. 
This can happen if the fields in the loop are gauge bosons; for minimally 
coupled matter the required vertex is forbidden by symmetry considerations.
Again, for $A_\mu^i$ with $(-,-)$ parity, no boundary counterterms exist and 
logarithmic divergences are absent.
When $A_\mu^i$ has parity $(+,+)$ and the fields in the loop are both odd, 
all divergences are canceled for $a=-b$; performing a Fourier decomposition 
in eq.(\ref{ct2}) only operators corresponding to transitions $2n\to 2n+4k+2$ 
are present. This means that the diagram of fig.(\ref{fig:diagrams}b) with odd 
modes in the loop and zero mode external lines do not give
rise to logarithmic divergences, as one can check by an explicit computation.
This explains why in the matching functions (see the following section) all the 
$\mu$-dependence comes from even modes.

\section{The matching equation}

Our goal is to relate the $SU(5)$ coupling $g_5$ to the 
$SU(3)\times SU(2)\times U(1)$
couplings measured at the weak scale.
In our scenario two different energy scales appear: the weak scale $M_Z$ and 
$1/R\gg M_Z$ which sets the typical mass for the heavy modes.
A useful way to deal with very heavy gauge fields was outlined by Weinberg 
in \cite{weinberg}. The idea is to construct a low energy effective 
$SU(3)\times SU(2)\times U(1)$ gauge
theory containing only light fields integrating out the heavy particles.
The matching consists in relating $g_5$ to the $SU(3)\times SU(2)\times U(1)$ 
couplings $g_i$ at the matching scale $\mu$. 
The effective action $S_{\text{eff}}$ is defined by
\begin{equation}
e^{i S_{\text{eff}}[\varphi]} = \int D\Phi \, e^{i S [\varphi,\Phi] +
 i S_{\text{GF}}[\varphi,\Phi]} \, .
\label{effective}
\end{equation}
Here the heavy fields are all the massive KK modes
$\Phi=(V^{(n)}, \Sigma^{(n)}, H^{(n)},\bar H^{(n)},H^{(n)}_c,\bar H^{(n)}_c)$ 
and the effective action depends on the zero modes fields 
$\varphi=(V^{a\, (0)}, H_2^{(0)},\bar H_2^{(0)})$.

Once a suitable gauge fixing  term $S_{\text{GF}}$ has been added to the 
action, the integration over the heavy fields is well defined. 
Splitting the gauge field into 
a ``light'' $\al$ and an ``heavy'' $\ah$ part, $A_M = \al_M + \ah_M$ 
\begin{equation} \label{eq:hl}
\begin{split}
\al_M &= A^{a\, (0)}_M T^a \\
\ah_M &= \sum_{n=1}^\infty A^{a\, (n)}_M \, T^{a} \, \cos [2ny/R] +
 \sum_{n=0}^\infty A^{\hat a\, (n)}_M \, T^{\hat a} \, \cos [(2n+1)y/R] \,,
\end{split}
\end{equation}
the original action $S$ has the following background symmetry
\begin{equation}
\begin{gathered}
\delta \al_M = {\cal D}_M   \xi^{{}^L} \, , \quad \delta \ah_M = i 
[ \ah_M, \,   \xi^{{}^L} ]  \\ 
{\cal D}_M = \partial_M + i[\al_M,\,\cdot\,] \quad ,
\end{gathered}
\label{back}
\end{equation}
which coincides with  $SU(3)\times SU(2)\times U(1)$ gauge transformations;
${\cal D}_M$ is the covariant derivative  with respect to the background  
$\al$. It is convenient to choose a gauge fixing for the heavy modes
which respects the low energy gauge transformation of eq.(\ref{back}) \cite{weinberg}. 
This way $S_{eff}$ is gauge invariant and the effective gauge coupling
is read just by looking at the vector kinetic term, without further need to look at the 
three point vertex. The simplest choice is the unitary gauge
\begin{equation}
\partial_5 A^5 = 0
\label{eq:uni}
\end{equation}
in which there are no ghosts. Another possible choice is the following
background covariant 't Hooft $\xi$ gauge \cite{weinberg}
\begin{equation}
{\cal L}_{\text{GF}} = 
\sum_{n=1}^\infty \, - \frac{1}{2 \xi} \left( \partial^\mu
A_\mu^{(n) i} - \frac{g_5}{\sqrt{2 \pi R}} \, f^{ija} \, A_\mu^{(n) j} \,
 A_\mu^{(n) a}  - \xi \frac{in}{R} \, A_5^{(n)i} \right)^2 \quad ,
\label{eq:xig}
\end{equation}
where $A_5^{(n)}$ play the role of the Goldstone bosons.
The functional integration over the heavy fields gives a matching relation 
between the  running low energy coupling $g_i(\mu)$ and the high energy 
parameters
\begin{equation}
\frac{1}{g^2_i(\mu)} = \frac{2 \pi R}{g^2_5} + \Delta_i( \mu) + \lambda_i(\mu
R) \; .
\label{match}
\end{equation}
In eq.(\ref{match}) the first term is the  tree level contribution from
the 5-dimensional kinetic term;  the second one, $\Delta_i(\mu)$, also 
originates at tree level and represents the contribution of 
the 4D gauge kinetic operators localized on the orbifold fixed points;
finally, the matching functions
$\lambda_i(\mu R)$ encode the radiative contribution from the massive modes 
\cite{weinberg,hall}. As we already said, in dimensional regularization
we only get logarithmic divergences, so that the $\lambda_i$ depend 
logarithmically on $\mu$.
At one loop level, the diagram relevant for the matching 
function $\lambda_i$ is the one of fig.(\ref{fig:diagrams}b) with all
scalar, vector and fermionic massive fields circulating in the loop; it gives
in the dimensional reduction scheme with minimal subtraction ($\overline{\text{DR}}$)
\begin{equation}
\lambda_i(\mu R) = \frac{1}{96 \pi^2} \left \{ \big[b_i^S - 21 \, b_i^G + 8 \,  
b_i^F \big]
F_{e}(\mu R) + \big[\tilde{b}_i^S - 21 \, \tilde b_i^G + 8 \,  \tilde b_i^F
\big]
F_{o}  \right \} \; ,
\label{mfuncts}
\end{equation}
with 
\begin{equation}
\begin{gathered}
F_{e}(\mu R)  = {\cal I} -1 - \log (\pi) - \log (\mu R) \, , \qquad 
F_{o} = - \log 2 \, , \\[0.3cm]
{\cal I} = \frac{1}{2} \int_1^{+ \infty} dt \, \left(t^{-1} + t^{-1/2} \right) 
 \left[\theta_3(it) - 1 \right] \, , \qquad
\theta_3(it) = \sum_{n=-\infty}^{+\infty} e^{-\pi t n^2} \, .
\end{gathered}
\end{equation}  
We denote with $b_i^{S,G,F}$ the constants  $C_i(r)$ defined as
$\text{Tr} (T^a T^b) = C_i(r) \, \delta^{ab}$ in a representation $r$ of the SM
group $i=SU(3),SU(2),U(1)$  for real scalars (S), vector bosons (G) and Dirac
fermions (F) with ``even'' mass $2n/R$. The constants $\tilde b_i^{S,G,F}$ are the same
quantities for ``odd'' modes with mass $(2n+1)/R$.
A numerical integration gives ${\cal I}\simeq 0.02$.
The $\ln \mu$ dependence of the $\lambda_i$'s is due to
the even modes only ($F_{e}$), the diagram in fig.(\ref{fig:diagrams}b) with
odd modes circulating being finite. 
As discussed in section \ref{sect:boundary} heavy modes with definite parity
lead to a divergent term in diagram (\ref{fig:diagrams}b) which is $-1/2$ the
zero mode divergent contribution; as a consequence 
the coefficient of $\ln \mu$ in our matching functions is
$-1/2$ the one in the standard 4D case (see eq.(2) in \cite{hall}).
For the specific model considered we get 
\begin{equation}
\begin{gathered}
\frac{1}{g^2_i(\mu)} = \frac{2 \pi R}{g^2_5} + \Delta_i(\mu) + 
\frac{\beta_i^H}{8 \pi^2} \log (\mu R) + \frac{\beta_i^H}{8 \pi^2} 
 \big[\log  \pi + 1 - {\cal I} \big] + \frac{\tilde \beta_i^H}{8 \pi^2} \log 2
\\[0.7cm]
\beta_i^H = (3,\, 1, \, -3/5) \, ; \qquad  \tilde \beta_i^H = (1,\, 3, \, 23/5)
\qquad
 i=SU(3),SU(2),U(1)
\label{res}
\end{gathered}
\end{equation}
We explicitly checked that both choices (\ref{eq:uni}) and (\ref{eq:xig}) 
for the gauge fixing give the same result.
One can verify that the coefficient of $\log (\mu R)$ in (\ref{res}) agrees with the result
of ref. \cite{HN} after it is written at an arbitrary scale $\mu$. 

Eq.(\ref{mfuncts}) can be used to perform a next-to-leading order calculation
of gauge unification. In this equation the constant terms are specific of our chosen
scheme, while the dependence on $\mu R$ is universal. In a supersymmetric theory 
this dependence can be directly
understood in terms of rescaling anomalies of the low energy theory \cite{kaplu}.
Considering just one group factor and focusing on the $R$ dependence, eqs.(\ref{match},
\ref{mfuncts}) read
\begin{equation}
\frac{1}{g^2(\mu)}=\frac{2\pi R}{g_5^2}+\frac{1}{8\pi^2}\left(C(A)-\sum_r C(r)\right)\ln R+
\dots
\label{holomorphic}
\end{equation}
where by $C(A)$ we indicate the Casimir of the adjoint representation, the sum $\sum _r$
extends over the bulk matter multiplets which have a zero mode chiral superfield, and by the dots
we indicate the $R$ independent terms. The above equation can be interpreted as the
relation between the holomorphic (Wilsonian) coupling $g_5^2/2\pi R$ and the physical (1PI)
 $g^2$. Notice indeed that in the supergravity context the radius $R$ is part
of a chiral superfield, whose scalar component is $S=R+i B_5$, where $B_5$ is the
5th component of the graviphoton. By analyticity and by the fact that $B_5$ 
couples derivatively, the holomorphic coupling $1/g_h^2$ can only depend
linearly on $S$: $1/g_h^2+i\theta/32\pi^2=2\pi S/g_5^2 +{\rm const}$. Notice also
that in the low energy lagrangian the massless matter fields coming from the bulk
have a wave function $Z=2\pi R\propto S+S^\dagger$. Then one sees directly that
in eq.(\ref{holomorphic}), the term $\propto C(A)$ represents the well known rescaling
anomaly of the gauge multiplet wave function \cite{shifman,arkani} while the matter
contribution is just the Konishi anomaly \cite{konishi}. 

Taking the $\mu d/d\mu$ derivative on both sides of eq.(\ref{res}) we get
\begin{equation}
\mu\frac{d}{d\mu}\Delta_i= - \frac{\left(\beta_i^H+\bmssm_i\right)}{8\pi^2}
\label{runDelta}
\end{equation}
where $\bmssm_i=(-3,1,33/5)$ are the $\beta$-function coefficients in the MSSM.
The $\Delta_i$'s are free parameters and have in principle an important
impact on gauge unification. In order to preserve predictivity we need to be able
to control the $\Delta_i$ by reasonable assumptions. By eq.(\ref{runDelta}) it is clear
that it would certainly be unnatural to assume that the $\Delta_i$ are 
smaller than $O(1/8\pi^2)$. However it seems natural to assume $\Delta_i\sim O(1/8\pi^2)$
at some given scale $\Lambda$. Around $\Lambda$, the tree and quantum contribution
to $\Delta_i$ would be comparable, a sign that some strong dynamics happens at $\Lambda$.
Now, our 5D theory is strongly coupled at a scale $\sim 24\pi^3/g_5^2$, so it is natural
to assume that this very scale coincides with $\Lambda$. This set of assumptions 
forms the basis of naive dimensional analysis (NDA) (see for instance \cite{markus}). 
The resulting NDA estimate of
the $\Delta_i$ gives a relative correction $O(g^2/8\pi^2)$ to the gauge couplings at the 
unification scale. This effect is small and predictivity is preserved.
This situation is quite similar to that of ordinary 4D GUTs. In that case the uncertainties 
come from the GUT spectrum (see for instance \cite{BH}), 
which in turn depends on the superpotential mass parameters
and Yukawa couplings. The usual assumption is that all masses are roughly of the same
order and that Yukawas are of order 1. In this way there is no hierarchy in the spectrum,
threshold effects are small. Then the theory is predictive and most importantly
there is good agreement with the data. Like for the $\Delta_i$'s the standard assumption
on GUT masses and couplings is not unnatural. For instance, it is stable under RG flow.
But it remains just an assumption: any GUT model by itself could have a large
hierarchy of masses, precisely like we are familiar in the SM where the top and electron
mass differ by 6 orders of magnitude. Indeed it is more the other way around: the fact
that gauge unification works  well suggests that the GUT theory is weakly coupled with
a fairly non hierarchical spectrum.

Indeed, with some more careful thinking, our 5D model supplemented with the NDA
assumption seems slightly better on the predictive side. The point is that there
is a parametric separation between the strong dynamics scale and the matching scale $1/R$:
$\Lambda R\sim 8\pi^2/g_{GUT}^2>> 1$. If we consider eq.(\ref{res}) for $\mu=1/R$,
we have that the threshold correction is controlled by 
\begin{equation}
\Delta_i(1/R)=\Delta_i(\Lambda) -\frac{\left(\beta_i^H+\bmssm_i\right)}{8\pi^2}\ln (\Lambda R)
 \sim  -\frac{\left(\beta_i^H+\bmssm_i\right)}{8\pi^2}\ln (8\pi^2/g_{GUT}^2)
\label{nda}
\end{equation}
When $\Lambda R$ is very large the second term dominates the bare 
$\Delta_i(\Lambda)\sim 1/8\pi^2$, and represents a definitely calculable threshold correction
to gauge unification.  One can then check if this correction improves the agreement
with the measured value of $\alpha_s$. Of course, in doing so one should be
aware that in the realistic case $\ln 8\pi^2/g_{GUT}^2\sim 5$, which is not much 
bigger than 1:  the unknown
contribution $\Delta_i(\Lambda)$ may not be so negligible compared to the effect we 
are considering.
Notice that concentrating on just the $\ln\Lambda R$ term is practically equivalent to what
done in ref. \cite{HN}, where it was (implicitly) assumed that the $\Delta_i$
unify at the scale $\Lambda$. Our assumption is however justified in a different way.

\section{Phenomenology}

\begin{figure}
\centering
\epsfig{file=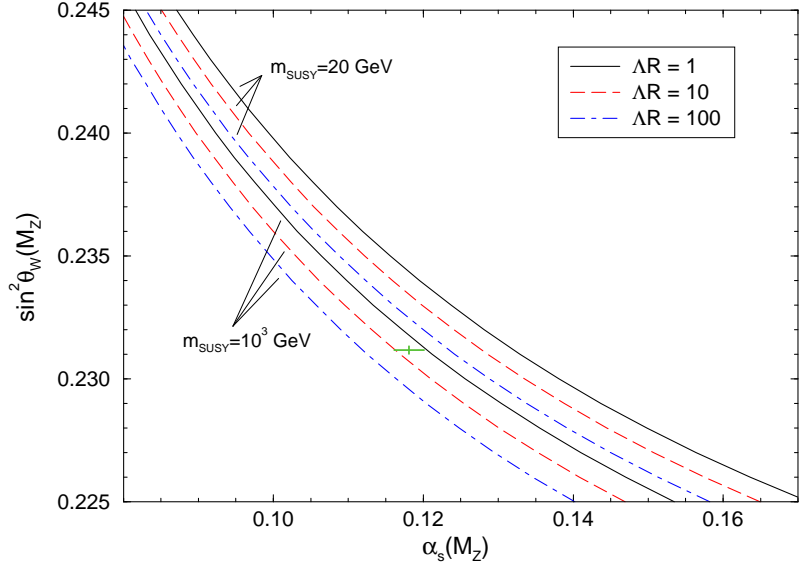,width=0.7 \linewidth}
\caption{{\it \small $\alpha_s(M_Z)$ as a function of $\sin^2 \theta_W(M_Z)$ 
varying $\ms$ in the range $20-10^3$ GeV for $\Lambda R=1,10,100$.
The experimental point is shown in green.}}
\label{assin}
\end{figure}

\begin{figure}
\centering
\epsfig{file=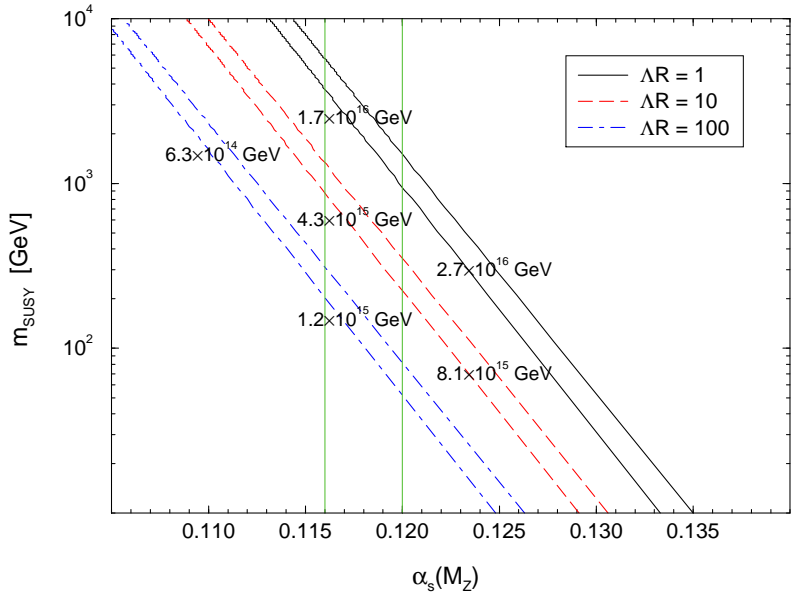,width=0.7 \linewidth}
\caption{{\it \small $\alpha_s(M_Z)$ as a function of $\ms$ for $\Lambda R=1,10,100$.
The region between lines of the same style represents the uncertainty coming from the
experimental error on $\sin^2 \theta_W(M_Z)$.
The region between the green vertical lines is the measured value of $\alpha_s(M_Z)$. 
The predicted values for $1/R$ are shown in the different regions.}}
\label{asmsusy}
\end{figure}

To compare with experiments we need to go down from the matching scale $\mu
\sim 1/R$ to the weak scale $M_Z$ solving the RG group equations at the 
next to leading order (NLO) with the initial conditions provided by the
matching equation. The running from 
$\mu$ down to $\ms$ is  determined by the MSSM spectrum; SUSY thresholds
are parameterized in terms of a single scale $\ms$ \cite{Ca}. From $\ms$ down to
$M_Z$ the running is the one
of the SM. We want to see the consequences on gauge unification of neglecting
$\Delta_i(\Lambda)$ in eq.(\ref{nda}). Though, as we argued, this neglect is motivated
(by NDA) only when $\Lambda R\sim 8\pi^2/g_{GUT}^2\sim 100$, we will remain general
and consider also smaller values of $\Lambda R$.   
Following \cite{Lan,hall,Jones} one has at~NLO
\begin{equation}
\begin{split}
\frac{1}{\alpha_i(M_Z^2)} =& \frac{4\pi}{g_{GUT}^2}  + \lambda_i(\Lambda R) +
\lambda_i^{\text{conv}} + \frac{\bmssm_i}{2 \pi} \log \frac{\Lambda}{M_Z} 
+ \frac{(\bsm_i - \bmssm_i)} {2 \pi} \log \frac{\ms}{M_Z}  + \\[0.3cm]
 & \frac{1}{4 \pi} \sum_j \, \frac{\bmssm_{ij}}{\bmssm_j} \,
\log\left[ 1 + \bmssm_j \alpha_G  \log (\Lambda /\ms )\right] \quad ; 
\end{split}
\label{run}
\end{equation}                            
$\lambda_i^{\text{conv}}$ is a conversion term from the $\overline{\text{DR}}$ 
scheme in which our computation has been
done to the $\overline{\text{MS}}$ scheme in which the SM $\alpha_i$ are defined 
\cite{Lan}. In practice we have used eq.(\ref{res}) with $\mu=\Lambda$ and imposed
$\Delta_i(\Lambda)=0$.
In fig.(\ref{assin}) we  plot  $\sin^2 \theta_W(M_Z)$ as a function of 
$\alpha_s(M_Z)$, varying $\ms$ in the range $20-10^3$ GeV for $\Lambda R=1,10,100$.
Increasing the value of $\Lambda R$, the curve goes in the right 
direction, approaching the experimental values  $\alpha_s(M_Z)= 0.1181 
\pm 0.002$, $\sin^2 \theta_W(M_Z) = 0.23117 \pm 0.00016$  
($\overline{\text{MS}}$ scheme) \cite{pdf}. Similar considerations apply to the
plot in fig.(\ref{asmsusy}) in which $\alpha_s(M_Z)$ and the unification scale $1/R$
are extracted using the experimental value of $\sin^2 \theta_W(M_Z)$. 
Indeed for the preferred value $\Lambda R\sim 100$ the band of prediction (fig. 2) 
sits precisely
on top of the measured values, with an interesting improvement over tree level
matching.  Unfortunately for this value of $\Lambda R$ one gets also $1/R\sim 10^{15}$
GeV, which is somewhat smaller that the lower bound from proton decay $1/R> 5 \times 10^{15}$
GeV \cite{HN}.  Indeed for $\Lambda R> 10$,  the right value of $\alpha_S$ 
is obtained only for $1/R<5 \times 10^{15}$ GeV. This is in agreement with what found in 
ref. \cite{HN}.
While the strict NDA assumption does not work
too well, we should be aware that we are talking about small effects:
$\ln \Lambda R\sim 5$ is not a very big number and the $\Delta_i(\Lambda)$ may play role. 
In the end, we are forced to conclude that even though the running of boundary
couplings gives an effect that goes in the right direction we still need the help
of the almost comparable initial value $\Delta_i(\Lambda)$ to agree with the data and satisfy
proton decay constraints.

\section{Conclusions}

We have applied the running and matching technique to study gauge unification
with one extra compact dimension. We have used dimensional regularization for which power 
divergences are absent and which is therefore very convenient for doing 
effective field theory studies \cite{weinberg}. Moreover dimensional reduction
is needed to consistently perform NLO calculations in supersymmetric theories. 
In orbifold models the radius dependence of the matching function is determined
by the mismatch between the RG evolution of the low energy $1/g_i^2$ and the RG
evolution of some gauge kinetic terms localized at the fixed points.  The bulk gauge
coupling does not run, so it is just a spectator in the matching. The main role
in determining corrections to gauge unification is played by the boundary kinetic
terms: these are free parameters and in general are not unified. 
Focusing on the $SU(5)$ model on
$S_1/(Z_2\times Z_2^\prime)$ of ref. \cite{kawa2,HN}, we have discussed to what extent strong
coupling assumptions based on NDA can control these unknown effects and lead to
a more predictive set up. Our conclusion is that these models, when supplemented with NDA,
stay more or less on the same  level as usual GUTs, where threshold effects are controlled
by assuming a non hierarchical GUT spectrum. Also in our case, in order to get a
 better agreement with the measured value of $\alpha_s$, one has to rely on small incalculable
effects.
\vspace{-0.2cm}

\section*{Acknowledgments}

We thank R. Barbieri, A. Strumia and F. Zwirner for several useful discussions and suggestions.
This work is partially supported by the EC under TMR contract
HPRN-CT-2000-00148.

\end{document}